# Differences among preparatory processes preceding major earthquakes in Kamchatka, 2014–2025, including the $M_W$8.8 event

Stanisław Lasocki[1], Beata Orlecka-Sikora[1]*, Mateusz Olszewski[1]

[1]Institute of Geophysics, Polish Academy of Sciences, Warsaw, Poland

*Beata Orlecka-Sikora orlecka@igf.edu.pl

## Abstract

Earthquake preparation involves progressive damage accumulation, strain localization, and changes in the organization of seismicity. We investigate preparatory processes preceding seven episodes of major earthquakes in the Kuril–Kamchatka subduction zone during 2014–2025, including the 29 July 2025 $M_W$8.8 Kamchatka megathrust earthquake. We track damage evolution using the seismic regularity coefficient dc and seismic strain dynamics coefficient SSD, defined for seismic series. Before 2022, each episode of strong seismicity was preceded by a pronounced dc minimum, commonly accompanied by an SSD minimum, followed by an increase in dc toward a maximum near the strong event or event group. After 2022, dc minima became shallower and were superimposed on a sustained upward trend. We interpret this transition as evidence that the metrics recorded two superimposed processes. The cycles before 2022 reflect complete strain-localization histories on southern or more distant structures that did not contribute to the segment ruptured in 2025. In contrast, the upward dc trend from 2022, with only shallow minima preceding Episodes 6 and 7, marks coupled behavior within the eventual main rupture segment, also influenced by stress transfer from southern source regions.

## Introduction

On 29 July 2025, a $M_W$8.8 megathrust earthquake occurred off the eastern coast of the Kamchatka Peninsula within the Kuril-Kamchatka subduction zone. The event ruptured a 500-km-long section of the plate interface, making it the largest earthquake recorded globally since the $M_W$9.1 Tohoku-oki earthquake of 2011 [1].

Various studies, including field observations, theoretical models, and laboratory experiments, have indicated that earthquake nucleation is a multiscale process that involves progressive damage accumulation, strain localization, and heterogeneous stress distribution [2–8]. Prior to a large failure, fault systems exhibit preparatory activity. This preparatory period involves the focusing of microscale damage, manifested as foreshocks and transient slip, which collectively signal the impending transition to rapid fault rupture [9].

Lasocki and Orlecka-Sikora introduced two metrics, defined for earthquake series, to track the evolution of the damage process leading to a major failure. These are the seismic regularity coefficient, *dc*, and the seismic strain dynamics coefficient, *SSD* (see the Method section). Temporal variations in these metrics are interpreted using the conceptual model of Orlecka-Sikora et al. [10,11], which focuses on the role of strain localization in the earthquake preparatory process (EPP) and its expression in the regularity of the seismic process. The *dc-SSD* metrics have been successfully used to track the EPP preceding stronger earthquakes across a range of magnitudes [10–14]. They are also currently used to monitor, in real time, the evolution of the damage process in the Colima region, which is part of the Mexican subduction zone [15].

In this study, we applied the *dc-SSD* metrics to decipher the characteristics of the EPPs preceding major earthquakes in the region of the great 2025 Kamchatka earthquake during 2014–2025. The analysis revealed differences in precursory patterns before major events, which we interpret using a continuum damage-mechanics description of progressive fault preparation (Orlecka-Sikora et al., GJI-26-0445.R1).

## Materials and Methods

### Seismotectonic setting

The Kamchatka Peninsula is located in the northeastern part of Asia, extending into the northwest Pacific between the Sea of Okhotsk and the Pacific Ocean and continuing southward into the Kuril Islands arc. The 29 July 2025 $M_W$8.8 earthquake occurred offshore along the Kuril-Kamchatka Trench, which marks the eastern margin of the peninsula. The event nucleated at 52.495°N, 160.240°E, with a focal depth of 35 km near Petropavlovsk-Kamchatsky [16]. It ruptured approximately 500 km southwestward, broadly parallel to the Kuril-Kamchatka Trench, with only approximately 100 km of the rupture extending northeastward [17]. This pronounced asymmetry has been interpreted as evidence for a predominantly unilateral and relatively simple rupture process [1]. The location and rupture geometry of the event are closely related to the subduction zone underlying Kamchatka and the Kuril Islands.

The Kuril-Kamchatka subduction zone forms part of an interface between the Pacific plate and the Okhotsk plate [18]. The Pacific plate subducts beneath the Okhotsk plate in a northwestward direction, broadly perpendicular to the Kuril-Kamchatka Trench. The convergence rate varies from 85 mm/y at 48°N to 75 mm/y at 55°N, approaching the Kamchatka-Aleutian triple junction to the north [19]. The subducting slab generally has a dip of approximately 55°, becoming shallower north of 55°N, where the dip decreases to approximately 35° [20]. The seismotectonic complexity also increases to the northeast, as the Obruchev Rise meets the subduction front [21]. The complex faulting pattern to the northeast is related to the nearly perpendicular junction of the Kuril-Kamchatka subduction zone with the Komandorskyi Sliver, which forms part of the Aleutian arc, and to the Emperor-Meiji seamount chain and the Kruzenstern fracture zone [22].

The area of the 2025 rupture overlaps substantially with the rupture zone of the 4 November 1952 $M_W$ 9.0 Kamchatka earthquake, with a similar segment affected by a great earthquake again after approximately 73 years [23–26]. The 1952 earthquake occurred at a depth of approximately 30 km and generated a destructive trans-Pacific tsunami. This historical earthquake is interpreted as a unilateral rupture with a relatively smooth slip distribution [27], similar to the 2025 event.

Based on the spatial extent of the 2025 seismicity and its relationship to the 1952 rupture zone, a segment of the Kuril-Kamchatka subduction zone was selected as the study area (Figs. 1 and 2). The area encompasses the principal seismicity associated with the 2025 earthquake while extending sufficiently to include events at depths of up to 100 km. The southern boundary was placed at latitude 48°N, corresponding to the southern extent of the 2025 seismicity and the limit of the seismicity catalog provided by KB GS RAS [28]. The northern boundary was placed across the trench because, beyond 55°N, the slab geometry and the seismotectonic setting change. The remaining boundaries encompass activity along the subduction interface and run parallel to the Kuril-Kamchatka Trench. The polygon was therefore designed to isolate a megathrust segment with a relatively uniform tectonic setting, while excluding adjacent tectonically distinct regions (Fig. 2). Exact coordinates of the bounding polygon are presented in the Data preparation section.

The decade preceding the 2025 $M_W$8.8 Kamchatka earthquake was characterized by persistent seismic activity across the region (Fig. 1). The earthquakes considered here include events within the research area at depths up to 100 km, as well as deeper and more distant regional earthquakes. The latter are included because large earthquakes outside the selected area may influence the regional stress field and seismicity beyond their immediate rupture zones [29,30]. The seismicity within the research area is heterogeneous, with two distinct concentrations of predominantly thrust activity that emerged over the study period.

The earlier concentration is located in the southern part of the research area, whereas later activity became concentrated farther north, close to the epicenter of the eventual 2025 event. We identified seven episodes of strong earthquakes during the ten-year period preceding the 2025 megathrust.

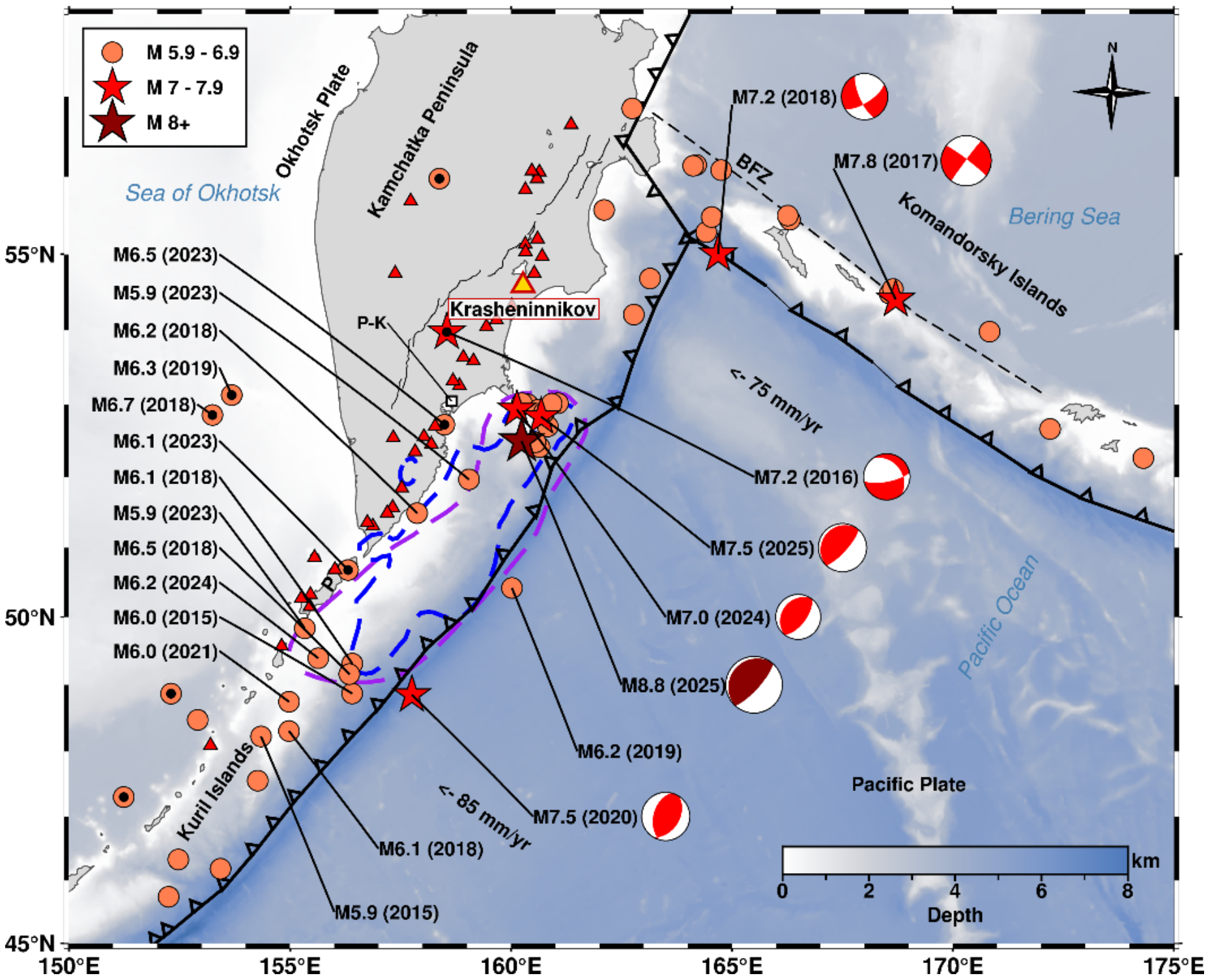

*Figure 1. Map of the vicinity of Kamchatka Peninsula showing the spatial distribution of seismic and tectonic features. Major and minor tectonic features from the GEM Global Active Faults Database [62] are represented by thick and thin black lines, respectively. Arrows indicate the direction and velocity of motion of the Pacific Plate relative to the Okhotsk plate [19]. Active volcanoes from the Global Volcanism Program [63] are represented by triangles, with Krasheninnikov volcano highlighted. The area that experienced at least 2 m of slip during the 2025 $M_w$8.8 rupture is outlined by a blue dashed line based on Liu et al. [1], with adjacent $M_w$5.9+ earthquakes labeled. The purple outline represents the rupture area of the $M_w$9.0 1952 earthquake from MacInnes et al. [25]. A thin dashed line represents the Bering Fracture Zone (BFZ). The city of Petropavlovsk-Kamchatsky is marked by a square (P-K). P marks Paramushir Island. Notable earthquakes (MW5.9+) that occurred between 1 January 2015 and the $M_w$8.8 Kamchatka earthquake on 29 July 2025 are shown by markers; stars denote magnitudes of 7 or greater, and dots denote events deeper than 100 km. Earthquakes of magnitude 5.9 or greater surrounding and downdip of the rupture area are labeled. The labels of the $M_w$7+ earthquakes are accompanied by their moment tensor solutions from GCMT [34,35], with red used for magnitudes from 7 to 8 and dark red for magnitudes above 8.*

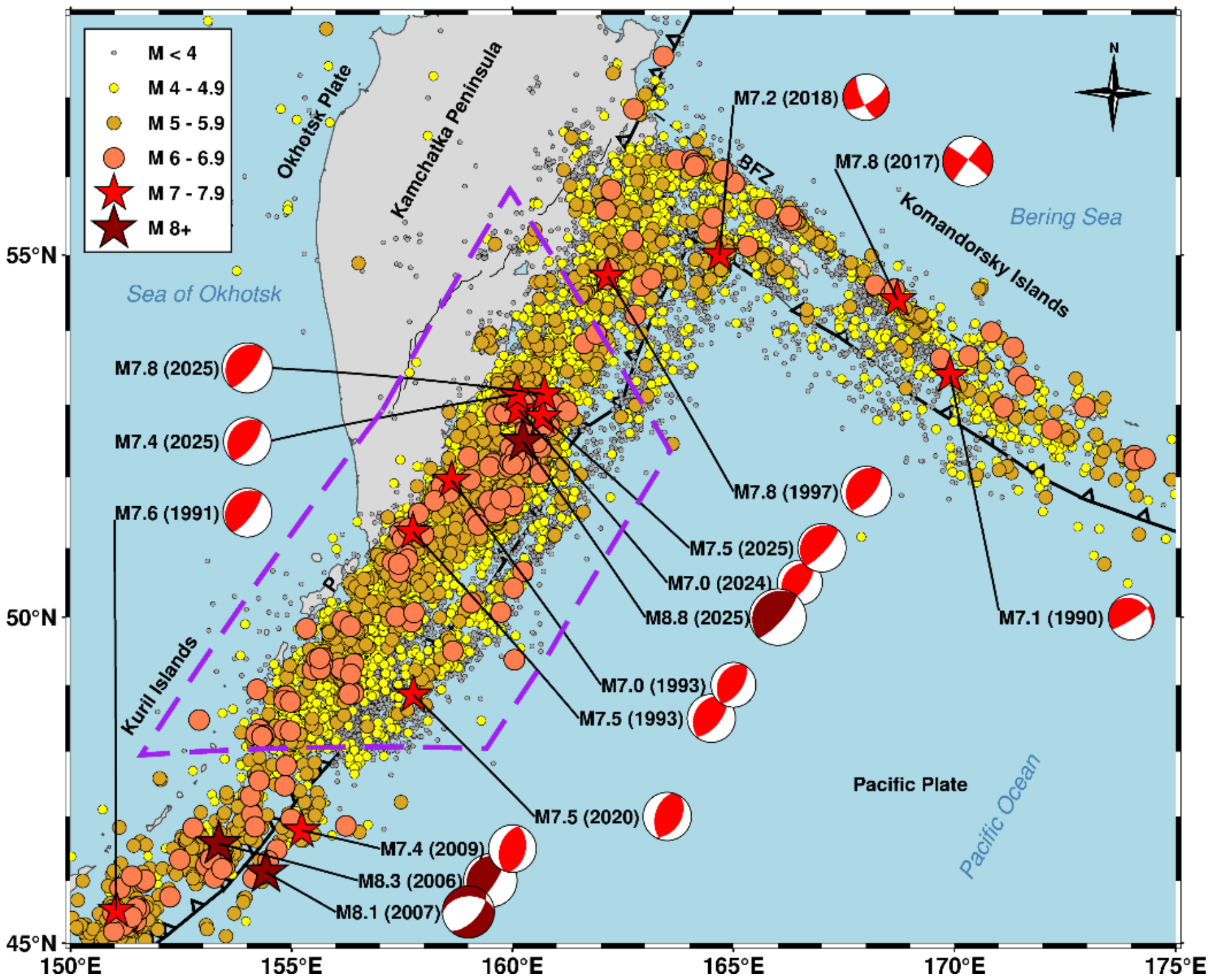

*Figure 2. Map of the vicinity of Kamchatka Peninsula showing the spatial distribution of seismicity and tectonic features. Major and minor tectonic features from the GEM Global Active Faults Database [62] are represented by thick and thin black lines, respectively. A thin dashed line represents the Bering Fracture Zone (BFZ). The city of Petropavlovsk-Kamchatsky is marked by a square (P-K). P marks Paramushir Island. Seismicity between 1 January 1990 and 5 April 2026 at depths of up to 100 km is shown by markers, with stars denoting $M_W$7+ events. The labels of the strongest earthquakes are accompanied by their moment tensor solutions from GCMT [34,35], with red used for magnitudes from 7 to 8 and dark red for magnitudes above 8. The violet outline indicates the area from which data were selected for analysis.*

Episode 1 comprises significant earthquakes that occurred from 2015 to 2016. On 14 October 2015, an $M_W$6.0 earthquake occurred approximately 150 km southeast of Paramushir Island at a depth of 12 km, followed by an $M_W$5.9 event on 16 November near 48.3°N at a greater depth of 35 km. The activity in this period was not confined to the shallower section of the subduction zone. On 30 January 2016, an $M_W$7.2 oblique-normal earthquake occurred approximately 100 km north of Petropavlovsk-Kamchatsky at a depth of 177 km [31]. Although this event occurred outside the study area, its magnitude and proximity to the epicenter of the 2025 mainshock justify its inclusion in the analysis of seismicity.

Episode 2 comprises a larger number of moderate-to-strong earthquakes that occurred between 2018 and 2019, including both shallow seismicity and the deepest earthquake of the entire decade. On 6 July 2018, an $M_W$6.2 earthquake occurred at a depth of approximately 45 km, followed by an $M_W$6.1 event on 10 August. A further sequence occurred in October, beginning with an $M_W$6.1 thrust earthquake on 9 October at a depth of approximately 19 km. The following day, an $M_W$6.5 earthquake occurred in the same area and at a similar depth. This group of earthquakes was followed by two strong events in neighboring sections of the subduction zone. On 20 November, an $M_W$6.7 strike-slip earthquake occurred within the subducting slab at a depth of approximately 460 km. The second, an $M_W$7.2 earthquake, occurred approximately 200 km north of the study area's northern boundary at a depth of 17 km on 20 December and was associated with oblique faulting near the junction between the Kamchatka and Aleutian island arcs. The last significant event in Episode 2 was an $M_W$6.2 normal faulting earthquake that occurred on the oceanward side of the Kuril-Kamchatka Trench at a depth of 9 km on 29 March 2019.

The largest earthquake within the earlier, southern concentration was an $M_W$7.5 event that occurred oceanward of Paramushir Island, close to the Kuril-Kamchatka Trench at a depth of approximately 58 km on 25 March 2020 (Episode 3). The event was located within the subducting Pacific Plate and has been interpreted as an intraplate, near-trench compressional earthquake associated with bending of the Pacific Plate [32]. Its depth and scale differ from the previous thrust earthquakes recorded in the surrounding area, although its location places it within the same broader concentration of seismicity.

Episode 4 comprises an $M_W$6.0 earthquake that occurred on 24 August 2021 near 48.75°N at a depth of 35 km, within the area that hosted the preceding cluster of major and strong earthquakes. It represents the continuation of notable seismic activity in the southern part of the research area. No further earthquake of comparable magnitude was recorded in that area during 2021 or 2022; strong seismicity became concentrated farther north in subsequent years.

The next group of notable earthquakes occurred in 2023 (Episode 5), with most events at intermediate depths along the coasts of the Kamchatka Peninsula and the Kuril Islands. On 3 April, an $M_W$6.5 earthquake occurred approximately 60 km south of Petropavlovsk-Kamchatsky at a depth of approximately 101 km. It was followed four days later by an $M_W$5.9 event near the southern coast of Paramushir Island at a depth of approximately 85 km. An $M_W$6.1 earthquake occurred beneath Shumshu Island, adjacent to the northern coast of Paramushir, at a depth of approximately 137 km. The last event in this group was an $M_W$5.9 earthquake near 52°N on 23 December. The distribution of these earthquakes outlines the northwestern, downdip side of the highly active portion of the subduction zone relative to the rupture area of the great 2025 event.

A greater concentration of significant earthquakes occurred farther north in 2024 (Episode 6). On 17 August, an $M_W$7.0 earthquake occurred 100 km east of Petropavlovsk-Kamchatsky at a depth of approximately 29 km. Two subsequent earthquakes were located less than 20 km from the previous event at a similar depth. The first, an $M_W$6.0 event, occurred on 30 August and was followed by an $M_W$5.9 earthquake exactly one month later. The activity was not confined to this cluster, as the last earthquake in this period was an $M_W$6.2 event that occurred on 23 October south of Paramushir Island, in the southern part of the region, at a depth of 40 km. A further $M_W$5.9 earthquake occurred on 8 December at a depth of approximately 89 km near the southwestern end of the study area. Most of the activity in this group of earthquakes was therefore concentrated close to the future 2025 mainshock epicenter, and additional events occurred at the southern end of the eventual rupture area.

The final group of significant earthquakes occurred in July 2025, immediately preceding and including the 29 July $M_W$8.8 earthquake (Episode 7). On 20 July, an $M_W$6.7 earthquake occurred 30 km east of the 2024 cluster at a depth of approximately 23 km. It was followed approximately 20 minutes later by an $M_W$7.5 earthquake less than 30 km to the south at a depth of approximately 34 km. Two further earthquakes, of $M_W$6.7 and $M_W$6.6, occurred later the same day, followed by an $M_W$6.0 event on 22 July and by $M_W$6.1 and $M_W$6.0 earthquakes on 24 July. The events were shallow and occurred at depths from approximately 10 to 35 km, comparable to the portion of the subduction interface subsequently ruptured by the great earthquake. All these earthquakes were concentrated within 60 km of the eventual epicenter of the mainshock. Finally, the $M_W$8.8 mainshock occurred 121 hours (five days) after the last event on 24 July.

In summary, during the 10 years preceding the $M_W$8.8 earthquake, the spatial distribution of seismic activity changed progressively. In the earlier part of the decade, major events were concentrated in the south of the study area. After 2022, they became progressively concentrated farther north, initially including intermediate-depth earthquakes near the northwestern boundary of the area of interest. This change became particularly evident in 2024, when an $M_W$7.0 earthquake and subsequent events occurred within the area surrounding the epicenter of the 2025 mainshock. In July 2025, a concentrated sequence of strong, shallow earthquakes occurred during the nine days preceding the great rupture.

### Data preparation

The initial data set was downloaded from the International Seismological Centre (ISC) [33] and covered the period from 1 January 1990 to 5 April 2026. It consisted of two parts: the reviewed ISC Bulletin through 31 March 2024 and the unreviewed bulletin thereafter. The resulting catalog consisted of earthquakes within 45°–65°N and 150°–175°E, with focal depths of up to 100 km.

Earthquakes for which $M_W$ values were reported to the ISC by GCMT [34,35] or National Earthquake Information Center (NEIC) [36] formed the basis of the catalog. For events lacking $M_W$ values, we used local magnitudes ($M_l$) reported to the ISC by the Kamchatka

Branch of GS RAS. These $M_l$ values were then converted to $M_W$ using a relation from Abubakirov et al. [37]:

$$M_W = M_l - 0.40;\ 3.4 \le M_l \le 6.4 \qquad (1)$$

Earthquakes with converted magnitudes were then added to the catalog. The final compiled catalog contained 27,899 earthquakes within the magnitude range 3.0–8.8 (Fig. 2).

We sorted the catalog chronologically. The catalog was then spatially restricted to the selected study area, defined by the polygon in Table 1, resulting in a subset of 17,848 earthquakes.

**Table 1. Coordinates of the selected study area.**

| Latitude | Longitude |
|---|---|
| 52.9303 | 157.0655 |
| 55.8386 | 159.9448 |
| 52.3643 | 163.5666 |
| 49.4256 | 160.7902 |
| 48.0419 | 159.4090 |
| 47.9423 | 151.5955 |

Figure 3 shows the cumulative number of recorded earthquakes in the selected area from 1 January 1990 to 5 April 2026. The activity rate is not homogeneous. Four distinct periods are evident: 1990–1999, 2000–late 2013, 2014–July 2025, and July 2025–2026. Rapid increases in the number of events correspond to aftershock sequences of the strongest earthquakes. To ensure a uniform long-term activity rate, we restricted the catalog to the period from January 2014 to the last event before the onset of Episode 7 on 20 July 2025. The resulting catalog contains 5132 earthquakes in a time span of 11 years, 7 months, and 29 days. The activity in this period is relatively stable. Increased seismic activity preceding the main earthquake can be observed during the first seven months of 2025.

The magnitude of completeness of the final catalog, $M_c = 3.6$, was determined using the modified Goodness-of-Fit Test [38]. Only the complete portion of the catalog was used, yielding N = 2,682 events within the selected polygon and an average of 232.24 earthquakes per year. The spatial distribution of the final subset is shown in Figure 4.

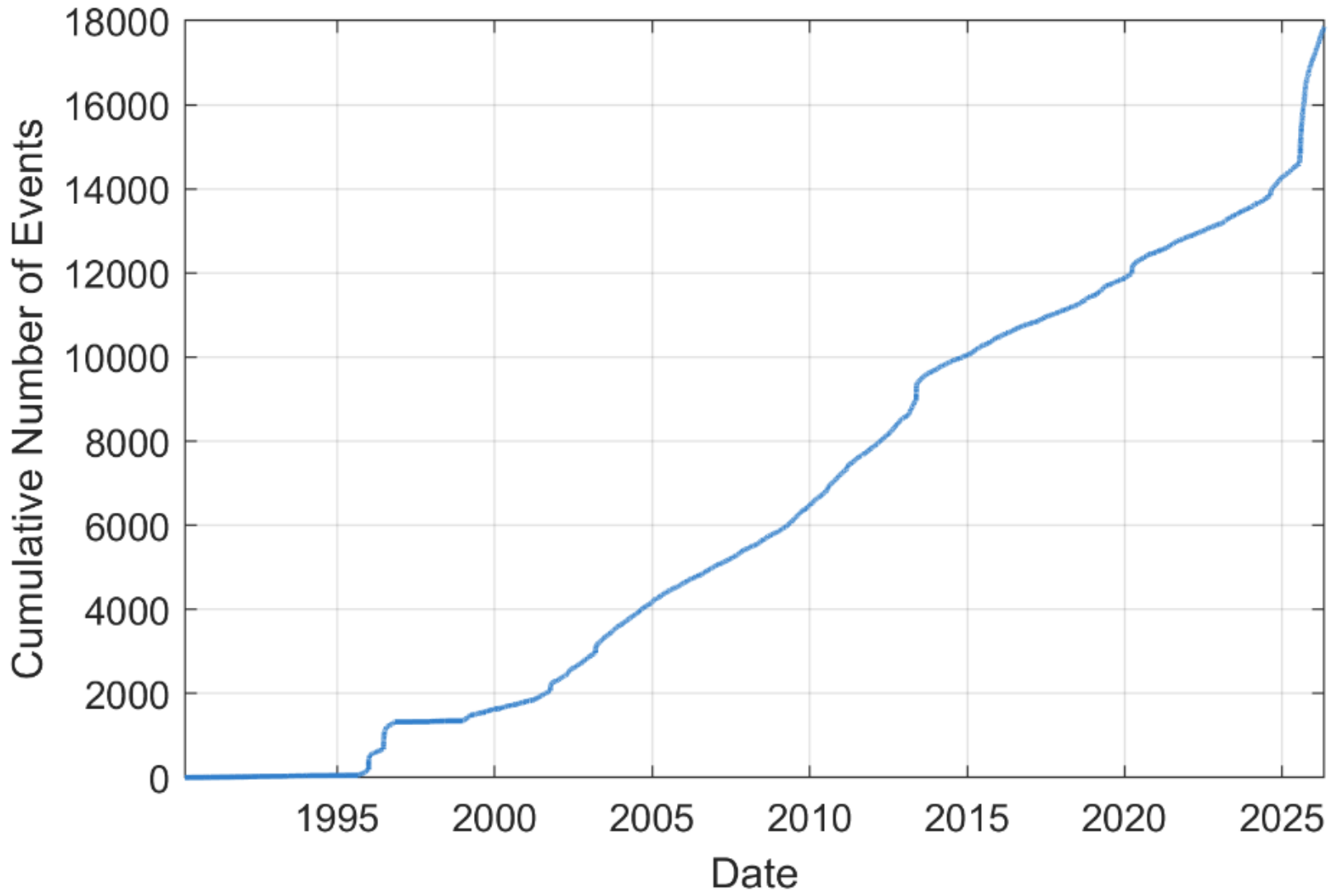

*Figure 3. Cumulative number of earthquakes in the extracted catalog from January 1990 to April 2026.*

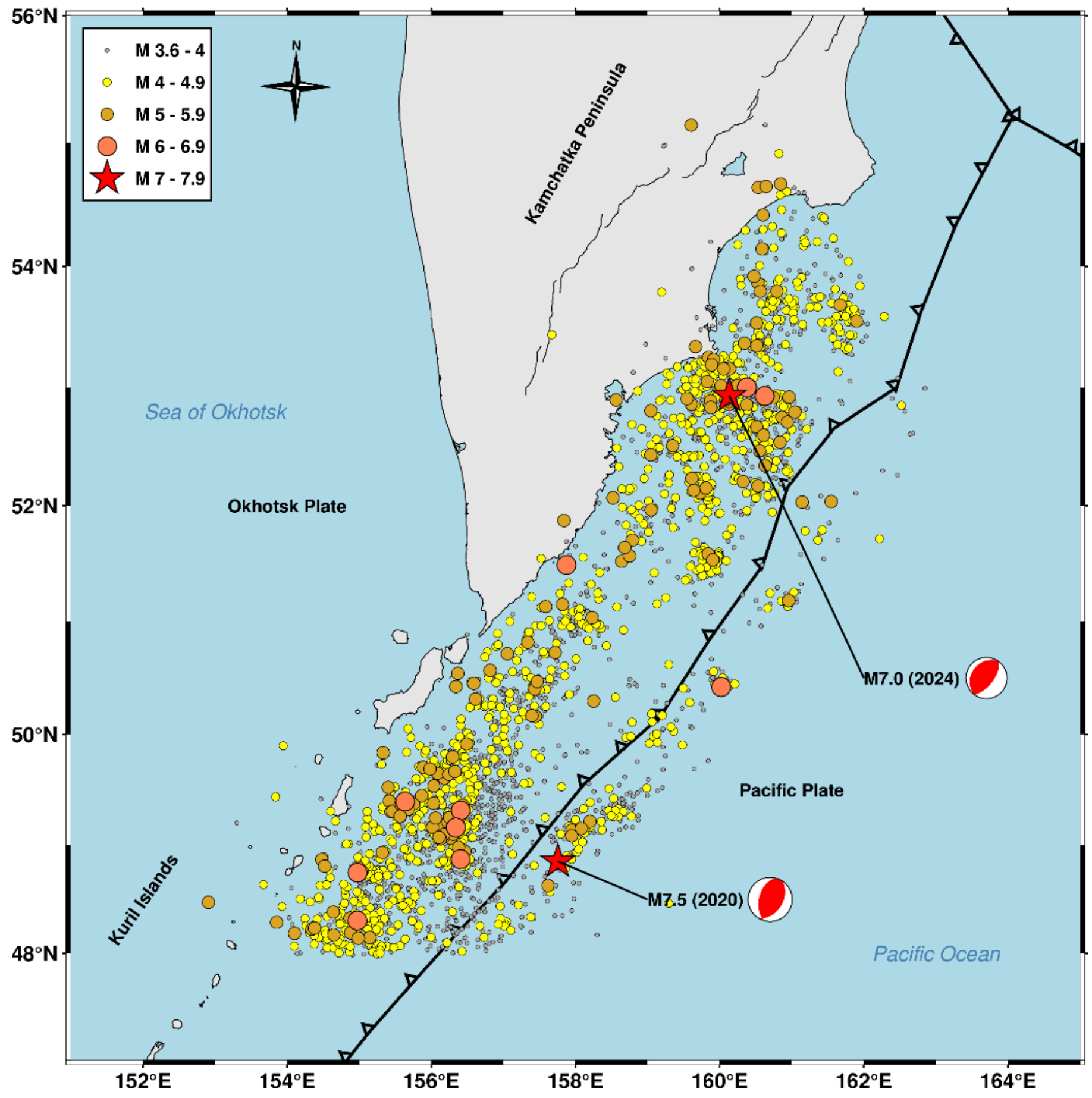

*Figure 4. Map of the vicinity of Kamchatka Peninsula showing the spatial distribution of tectonic features and the final earthquake set. Major and minor tectonic features from the GEM Global Active Faults Database [62] are represented by thick and thin black lines, respectively. Arrows indicate the direction and velocity of motion of the Pacific Plate relative to the Okhotsk plate [19]. Seismicity in the complete ($M_W$3.6+) data set from 1 January 2014 to 20 July 2025 is shown by markers, with stars denoting $M_W$7+ events. The labels of the strongest earthquakes are accompanied by their moment tensor solutions from GCMT [34,35], with red used for magnitudes from 7 to 8 and dark red for magnitudes above 8*

## Method

The method used in this study was previously applied to investigate other earthquake preparatory processes (EPPs) and is presented in detail in Lasocki et al. [12–14] and Orlecka-Sikora et al. [10,11]. Here, we provide only a concise overview.

The EPP is analyzed using two metrics. The first is the seismic regularity coefficient, *dc*, which is based on a three-dimensional representation of seismicity. Starting from an earthquake catalog, events above a prescribed magnitude threshold, no lower than the magnitude of completeness, are selected. Each selected event is characterized by the waiting time since its predecessor, the epicentral offset from that predecessor, and its magnitude, transformed into equivalent dimensions [39,40].

For a moving window of *K* events, *dc* is defined as the average pairwise distance between the transformed points:

$$dc = \frac{2}{(K-1)K}\sum_{k=1}^{K-1}\sum_{j=k+1}^{K}\sqrt{\left(DT_k - DT_j\right)^2 + \left(DR_k - DR_j\right)^2 + \left(MC_k - MC_j\right)^2} \quad (2)$$

Lower *dc* values indicate greater similarity among events in their waiting times, epicentral offsets, and magnitudes and, therefore, a more regular seismic process. Higher values indicate greater heterogeneity in the fracturing process.

The second metric, the seismic strain dynamics coefficient (*SSD*), is defined as:

$$SSD = log\left(\frac{\sum_{k=1}^{K} M_{0,k}^{1/3}}{\sum_{k=2}^{K} dt_k}\right) \quad (3)$$

where $M_{0,k}$ is the scalar seismic moment of the *k*-th event and $dt_k$ is the corresponding waiting time. If seismic moments are unavailable, they are approximated from magnitudes using the Hanks-Kanamori relation [41], even when the magnitudes are not moment magnitudes. In the current implementation, we use the equivalent dimensions of $M_{0,k}^{1/3}$ and $dt_k$ in equation (3). *SSD* obtained in this way retains the same relational characteristics as *SSD* calculated from the original $M_0$ and $dt$, but has better statistical properties.

*SSD* approximates the logarithm of the average rate of inelastic strain accumulation and tracks the rate of rupture-network development. Thus, *dc* characterizes the organization and variability of rupture increments, whereas *SSD* characterizes the rate at which deformation evolves. Each *dc* and *SSD* value is assigned to the occurrence time of the last event in the corresponding window.

The *dc* and *SSD* trajectories vary with the selected magnitude threshold, i.e., the magnitude threshold $M_{cut}$ above which events are selected to evaluate *dc* and *SSD*, although their main features, including trends and trend turning points, are generally preserved. Trajectories based on stronger earthquakes (high $M_{cut}$) do not capture processes that produce smaller events, whereas the inclusion of smaller events may introduce variability that obscures signals preceding stronger earthquakes. To reduce dependence on an arbitrarily selected threshold, we calculated the metric trajectories for a set of thresholds spaced by 0.1 magnitude units. For each threshold, we selected the number of events *K* in the moving window such that the windows for different $M_{cut}$ values had the same duration. We then aggregated the resulting trajectories by calculating a trimmed mean after removing the largest and smallest *dc*/*SSD* values at each time point *t*.

The seismic strain dynamics (*SSD*) and seismic regularity coefficient (*dc*) metrics are grounded in a continuum damage-mechanics description of progressive fault preparation, derived and validated across various scales by Orlecka-Sikora et al. (GJI-26-0445.R1). Under quasi-static loading, the local inelastic shear-strain rate $\dot{\gamma}_{in}(\boldsymbol{x},t)$ within a fault-zone volume follows a damage constitutive law (Eq. 1 in Orlecka-Sikora et al. (unpublished manuscript)) that extends the continuum damage rheology framework of Lyakhovsky et al. [42], further developed by Lyakhovsky and Ben-Zion [43], by incorporating a nonlocal, curvature-dependent stress-redistribution term [44,45]:

$$\dot{\gamma}_{in}(\boldsymbol{x},t) = \max\{\tau_{eff}(\boldsymbol{x},t)/\eta(D(\boldsymbol{x},t)) - \chi(D(\boldsymbol{x},t))\nabla^2\tau(\boldsymbol{x},t), 0\} \quad (4)$$

where $\boldsymbol{x}$ denotes spatial position, $t$ is time, $\nabla^2$ is the spatial Laplacian with respect to $\boldsymbol{x}$, $\tau_{eff}$ is the damage-reduced driving stress, $D \in [0,1]$ is the scalar damage variable, $\eta(D)$ is the damage-softened effective viscosity controlling the rate of local inelastic flow, and $\chi(D)$ is the redistribution mobility governing the nonlocal transfer of stress between neighboring regions. Orlecka-Sikora et al. (GJI-26-0445.R1) show that, when the spatial mean of Eq. (4) is taken over the actively deforming volume, the mean inelastic strain rate $\langle\dot{\gamma}_{in}\rangle$ increases monotonically as $\eta(D)$ declines under sustained driving stress. A second-moment analysis of the same relation, combined with a quasi-static elastic stress balance, shows that the spatial variance, $\mathrm{var}(\dot{\gamma}_{in})$ is instead controlled by the competition between the rate at which evolving damage generates new, spatially independent stress heterogeneity ($Q_\tau$) and the rate at which $\chi(D)$ damps existing heterogeneity through redistribution. As spatially competing damage clusters coalesce into a single dominant deformation zone, the production of independent heterogeneity declines; when this decline outpaces the damage-induced increase in the sensitivity of the strain-rate field to stress fluctuations, $\mathrm{var}(\dot{\gamma}_{in})$ passes through a transient minimum before recovering as rupture-scale complexity subsequently develops. The theory predicts that the variance minimum should precede the maximum of the mean, providing a mechanical criterion for distinguishing coherent shear localization from distributed acceleration of damage across multiple, non-coalescing zones.

Orlecka-Sikora et al. (GJI-26-0445.R1) derived the formal connection between *SSD* and *dc* and the statistical moments of the inelastic strain-rate field. *SSD* was shown to be a monotonic proxy for $\langle \dot{\gamma}_{in} \rangle$. Empirical scaling relations linking seismicity rate, interevent time, and magnitude to the local inelastic strain rate further show that *dc* is proportional to the standard deviation of $\dot{\gamma}_{in}$. Based on this, lower *dc* corresponds to more spatially organized, localized deformation. Therefore:

(1) *SSD* rises as damage accumulates.
(2) *The dc* reaches a minimum as competing deformation sites coalesce into a coherent shear zone.
(3) The *dc* minimum precedes the *SSD* maximum.
(4) *The dc* subsequently recovers as rupture-scale complexity develops while *SSD* plateaus or declines.

This *dc*-minimum-before-*SSD*-maximum sequence is the primary diagnostic of the transition from distributed deformation to coherent shear localization.

Of particular diagnostic value are temporally coincident minima in *dc* and *SSD*. According to the five-stage model of system-size damage proposed by Orlecka-Sikora et al. [11], these minima are associated with the volume-expansion point (VEP), which marks the transition from Stage III to Stage IV, that is, from stable to unstable crack propagation.

Shear localization is not, however, guaranteed to run to completion on every fault segment. When the effective stress acting on a given volume includes an externally imposed contribution from adjacent, mechanically coupled segments or creeping zones, the total effective stress can be decomposed as $\tau_{eff} = \tau_{int} + \tau_{ext}$, where $\tau_{int}$ is the stress generated internally within the volume by its own evolving damage field, and $\tau_{ext}$ is the additional stress transferred into the volume from neighboring, mechanically coupled structures. Under this decomposition, the equilibrium variance of the inelastic strain-rate field acquires an additional, externally sourced term, $\mathrm{var}(\tau_{ext})$, that prevents the variance from collapsing to the low values achievable on an isolated, geometrically simple fault (Orlecka-Sikora et al., GJI-26-0445.R1). The dimensionless ratio $\varphi = var(\tau_{ext})/var(\tau_{int})$ quantifies this effect: for $\varphi \ll 1$, the isolated-fault *dc*-*SSD* cycle described above applies, whereas for $\varphi \sim 1$—particularly when external loading preferentially stresses regions that have not yet accumulated significant internal damage, producing negative covariance between $\tau_{int}$ and $\tau_{ext}$—the *dc* minimum is expected to be systematically shallower or absent, because inter-segment stress redistribution continuously sustains spatial heterogeneity that an isolated shear zone would otherwise shed. This distinction is central to interpreting the decade of seismicity preceding the 2025 $M_W$8.8 Kamchatka earthquake.

## Results

Figs. 5 and 6 present time series of the *dc* and *SSD* metrics for the period from 25 July 2014 to 16 July 2025. These series end before the occurrence of the $M_W$8.8 Kamchatka mainshock and its three $M_W$6+ foreshocks, which took place between 20 and 29 July 2025. To obtain the results shown in Fig. 5, we used the raw catalog, to which we applied four magnitude thresholds $M_{cut}$ = 4.0–4.3 in increments of 0.1. We used a sliding window with a constant duration of 254 days, corresponding to 69, 53, 45, and 37 events for the successive magnitude cutoffs, respectively. We then aggregated the individual results for the magnitude cutoffs as described in the Method section.

The results in Fig. 6 were obtained by applying the same magnitude thresholds to the catalog that had previously been declustered using the Reasenberg algorithm [46]. The declustering reduced the size of the catalog by 21 percent. The same fixed-duration sliding window of approximately 254 days was used; for the successive magnitude thresholds, this resulted in 55, 43, 35, and 30 events within the window, respectively. The curves in Fig. 6 represent the aggregated results for the magnitude thresholds. Hereafter, for simplicity, we refer to *dc* and *SSD* trajectories obtained from undeclustered data as *dc_u* and *SSD_u,* respectively, and to *dc* and *SSD* trajectories obtained from declustered data as *dc_d* and *SSD_d,* respectively.

In general, *dc*/*SSD* trajectories obtained from raw (undeclustered) seismic series (*dc_u* / *SSD_u*) are our primary diagnostic tools. However, when a strong earthquake is followed by a series of tightly clustered strong aftershocks, such an event sequence distorts these trajectories. The distortion manifests as an immediate drop in *dc_u* and a sharp increase in *SSD_u*. In such distorted parts of *dc_u* / *SSD_u* trajectories the response of these metrics to the EPP development may not be identifiable. To avoid misinterpretations of *dc_u* / *SSD_u* trajectories due to such distortions, we also compute the trajectories from declustered seismic series and use them as auxiliary tools. If sections of the primary *dc_u* / *SSD_u* trajectories exhibit shapes characteristic of aftershock effects, we may base our interpretation of these sections on the secondary *dc_d* / *SSD_d* trajectories.

There are two reasons why we use *dc_d* / *SSD_d* trajectories derived from declustered catalogs only in exceptional cases. First, declustering is an automated procedure that removes a significant portion of the earthquake catalog. Declustering algorithms rely on various concepts for identifying aftershocks, yet none of them guarantee that the removed events are indeed aftershocks or that all aftershocks have been removed. A detailed discussion of this issue can be found in Zaliapin and Ben-Zion [47]. Second, the declustered seismic series is considerably reduced. This produces a comparable reduction in the number of events within each data window. Consequently, the metric estimates have higher uncertainty. In extreme, although not uncommon, cases, to evaluate the *dc_d* / *SSD_d* trajectories correctly, a window must be extended over a very long, and sometimes unacceptable, time interval.

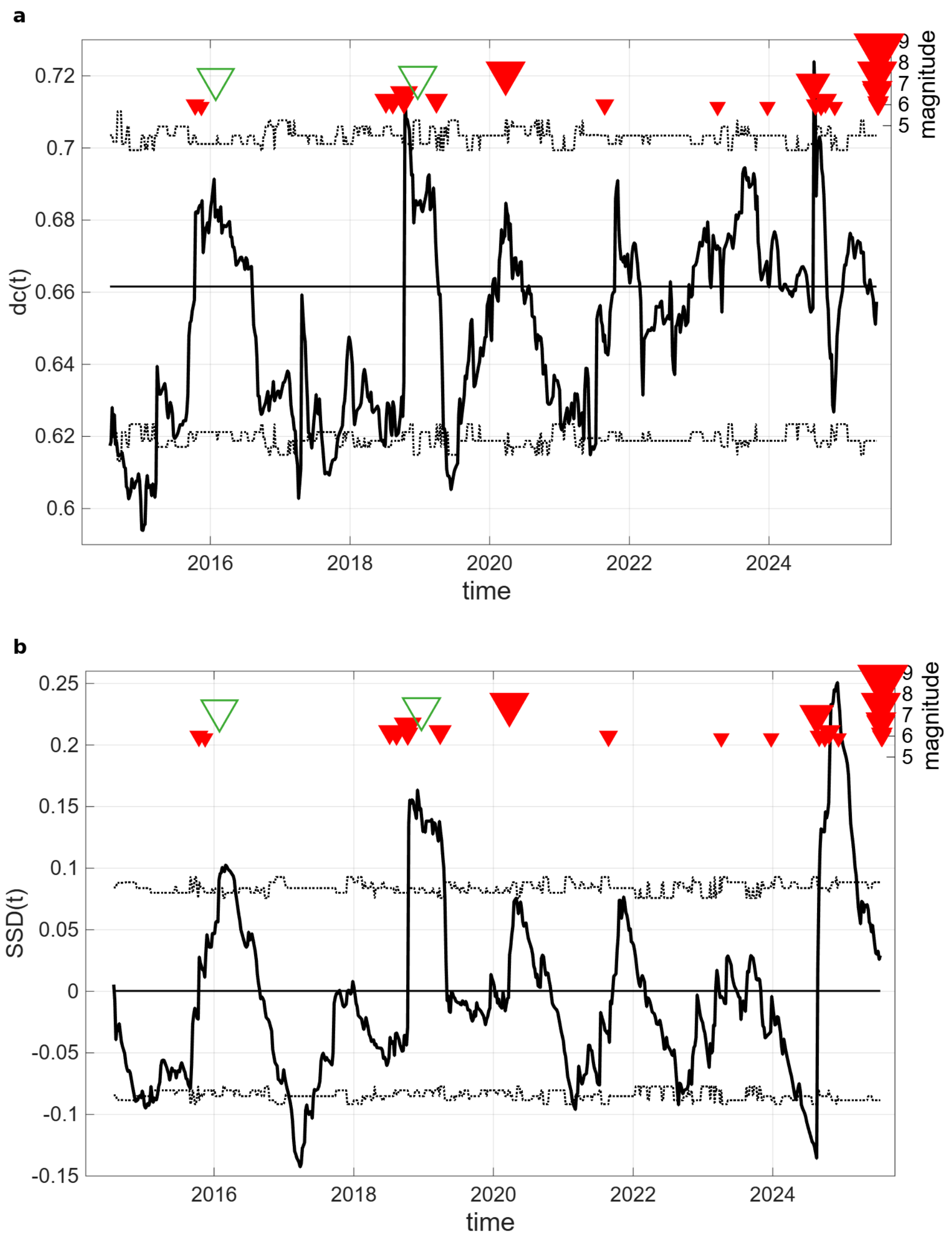

*Figure 5. Time series of (a) dc and (b) SSD from 25 July 2014 to 16 July 2025 computed from the raw catalog (heavy lines). Red triangles indicate the occurrence times of* $M_W$*5.9+ earthquakes and the triangle sizes are scaled by the earthquake magnitudes. Green triangles mark notable earthquakes in the region that are not included in the input dataset. Dotted lines are the bounds of the 90% confidence interval of dc/SSD for an acausal, random earthquake series (a random permutation of the input series); the black horizontal line is the expected value for the random case.*

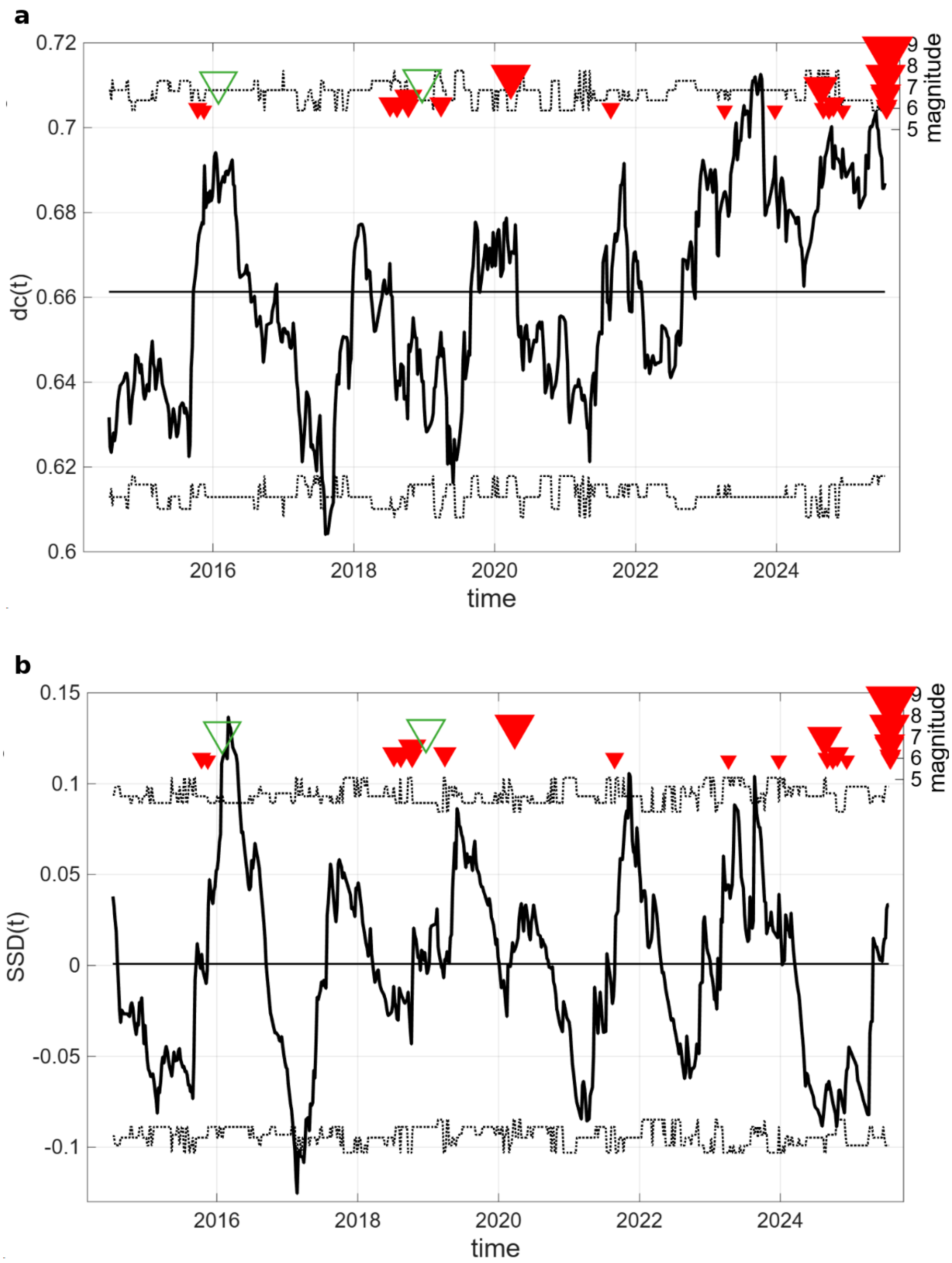

*Figure 6. Time series of (a) dc and (b) SSD from 25 July 2014 to 16 July 2025 computed from the declustered catalog (heavy lines). Red triangles indicate the occurrence times of $M_W$5.9+ earthquakes and the triangle sizes are scaled by the earthquake magnitudes. Green triangles mark notable earthquakes in the region that are not included in the input dataset. Dotted lines are the bounds of the 90% confidence interval of dc/SSD for an acausal, random earthquake series (a random permutation of the input series); the black horizontal line is the expected value for the random case.*

Comparing the trajectories in Fig. 5 with the trajectories in Fig. 6, one instance of the above-described distortion caused by a mainshock-aftershock sequence was identified. The distortion due to aftershocks following the $M_W$7.0 "Shipunsky" earthquake of 17 August 2024 complicated the interpretation of *dc_u*(*t*) and *SSD_u*(*t*) during the final period preceding the $M_W$8.8 mainshock. For this period, we therefore based our interpretation on the results from declustered data; in all other periods the trajectories from undeclustered data were interpreted.

Overall, until 2022, deep minima of *dc*(*t*) preceding stronger earthquake occurrences are seen in both Fig. 5a and Fig. 6a. These minima are followed by pronounced increases in *dc*(*t*), and the strong earthquakes occur near the maxima of *dc*. Furthermore, these minima of *dc*(*t*) coincide temporally with minima on the *SSD*(*t*) curves indicating volume-expansion points (VEPs), i.e., times of transition from stable to unstable crack propagation (from Stage III to Stage IV).

The earthquakes that we classified as Episode 1 are preceded in Fig. 5 by minima of *dc_u* and *SSD_u* around mid-January 2015. The $M_W$6.0 and $M_W$5.9 earthquakes located within the study area, and the $M_W$7.2 earthquake located outside this area, which followed these *dc_u* and *SSD_u* minima, occurred near the maxima of both trajectories. The lead times from the minima to the first pair of events and to the last earthquake in Episode 1 were approximately nine months and one year, respectively. The time variations of *dc_u* and *SSD_u* are well reproduced in *dc_d* and *SSD_d* trajectories (Fig. 6).

The *dc*(*t*) and *SSD*(*t*) trajectories preceding Episode 2 (February 2016–October 2018) are more complex. Between April and August 2017, *dc_u*(*t*) undergoes almost step-like changes (Fig. 5a). The deep *dc_u* minimum on 6 April 2017 coincides closely in time with an equally deep *SSD_u* minimum (Fig. 5b). Moreover, both minima are reproduced in the metric trajectories derived from the declustered data (Fig. 6). However, two additional *dc_u* minima occur before the Episode 2 earthquakes (Fig. 5a), on 14 September 2017 and 3 July 2018. The former corresponds to a very deep *dc_d* minimum, whereas the latter is correlated with a distinct *SSD_u* minimum. Lead times between these three *dc_u*(t) minima and the first earthquake of Episode 2 are ten months, five months, and only eight days, respectively. It is unclear which of these *dc_u* minima marks the volume-expansion point (VEP) preceding Episode 2; nevertheless, the fact that all three minima occurred before Episode 2 highlights the predictive potential of the *dc*/*SSD* metrics.

In contrast to the trajectories associated with Episode 2, those preceding Episode 3, the $M_W$7.5 earthquake, show a clear pattern. The *dc_u* trajectory steeply decreases and reaches a deep minimum on 12 June 2019. Subsequently, *dc_u* increases steadily to a maximum on 25 March 2020, the day of the $M_W$7.5 earthquake. The lead time between the *dc_u*(*t*) minimum and the $M_W$7.5 earthquake is approximately nine months. *SSD_u*(*t*) also decreases to a minimum in October 2019, four months after the *dc_u*(*t*) minimum. Minima preceding this earthquake are also visible on the *dc_d* and *SSD_d* trajectories.

Subsequent minima of *dc* and *SSD* precede the $M_W$6.0 earthquake on 24 August 2021 designated as Episode 4. The *dc_u*(*t*) trajectory reaches its minimum on 10 June 2021, two months before the $M_W$6.0 earthquake; *dc_d*(*t*) reaches its minimum on 7 May 2021, three months before the event. The *SSD* minima occur in March–April 2021. The mainshock occurs on the rising branches of all metric trajectories.

The pattern of the *dc*/*SSD* trajectories after 2022 differs from that observed before 2022. Distinct *dc* minima are still visible, but they are much shallower than those recorded before 2022 and appear only as local depressions superimposed on an overall increasing trend that continues until the mainshock. This upward trend is more clearly expressed in the *dc_d*(*t*) trajectory, because *dc_u*(*t*) is strongly disturbed during the final stage by aftershocks of the Episode 6 events.

The first *dc_u*(*t*) and *dc_d*(*t*) minima in 2022, occurring in March and August, respectively, and the subsequent increase in both trajectories can be associated with the two Episode 5 $M_W$5.9 earthquakes (7 April and 24 December 2023). These minima are temporally correlated with minima in *SSD_u*(*t*) and *SSD_d*(*t*).

From their maxima around August–September 2023, both *dc_u*(*t*) and *dc_d*(*t*) decline to distinct minima that are not particularly deep in absolute terms on 4 August and 22 May 2024, respectively. These minima correspond to *SSD_u*(*t*) and *SSD_d*(*t*) minima on 17 and 5 August 2024, respectively. We associate these minima with the Episode 6 earthquake group, which begins with the $M_W$7.0 “Shipunsky” earthquake of 17 August 2024. The interval between the VEP and this $M_W$7.0 event is between two weeks and three months.

From the occurrence of the $M_W$7.0 earthquake until the end of the study period, disturbances in the *dc_u*(*t*) and *SSD_u*(*t*) trajectories due to aftershocks prevent firm conclusions based on these metrics. During this period, *dc_d*(*t*) declines to a local minimum with a relatively high value of 0.68 on 5 March 2025. During this period, *SSD_d*(*t*) remains stable at low values until 3 April 2025. Based on the temporal coincidence of the *dc_d*(*t*) and *SSD_d*(*t*) minima, we infer that the VEP occurred in March 2025, approximately five months before the great $M_W$8.8 mainshock of 29 July 2025.

However, from its 5 March minimum, *dc_d* rises to a maximum on 25 May 2025 and then declines to another local minimum on 4 July 2025, 16 days before 20 July 2025, the date of the first two earthquakes of Episode 7. A minimum on 20 July 2025 is also discernible in the disturbed *dc_u*(*t*) trajectory. In contrast, *SSD_d*(*t*) increases markedly from 29 March 2025 until the end of the study period.

## Discussion

During 2014–2025, strong and major earthquakes occurred in the study region as two individual events and five groups, including the $M_W$8.8 great event. Each of these episodes of seismic activity was preceded by a minimum of the seismic regularity coefficient, *dc(t)*, which in most cases was temporally correlated with a minimum of the seismic strain dynamics coefficient, *SSD*(*t*). Thereafter, *dc(t)* increased, and the strong event or group occurred at or near the time of *dc* maximum. This result provides further confirmation of the previous conclusions [10–14] that the *d*c-*SSD* metric pair has high potential for earthquake forecasting.

Although the relational pattern "the *dc* minimum—a considerable increase in *dc(t)*—a strong earthquake" was repeated in all cases, the amplitudes of the associated *dc(t)* variations differed. Until 2022, the *dc* cycles—minimum, increase, maximum, and decrease—associated with strong earthquakes were clearly separated. The *dc* minima were well developed and deep, lying close to or below the lower 5% benchmark line. By contrast, after 2022, the *dc* minima were shallower, and the corresponding *dc* cycles were superimposed on a long-term upward trend of the *dc* trajectory. This effect is more clearly visible in the trajectory obtained from the declustered catalog because the rapid decrease of *dc(t)* from undeclustered data after the sixth episode of strong activity (August–December 2024), most likely associated with aftershocks, obscures the overall trend.

Interseismic GNSS coupling models show that the plate interface south of ~53°N—the segment that ruptured in both 1952 and 2025—undergoes essentially no slip to a depth of ~50 km, whereas the interface north of ~53°N accommodates substantial aseismic slip [48]. Bürgmann et al. [48] further show that models treating entire historic rupture zones as uniformly locked overpredict the observed kinematic coupling; they argue instead that great earthquakes nucleate on small, fully locked core asperities—on the order of 25% of the historic rupture area—embedded within a conditionally stable surrounding interface. A recent probabilistic, data-driven locking analysis [26] confirms and refines this picture: a principal locked cluster contains both the 1952 and 2025 nucleation points, and hierarchical clustering of the locking ensemble reveals internal structure within this cluster—five subclusters with mean locking probabilities ranging from ~0.98 to ~0.73—indicating that the "fully locked" segment is itself mechanically heterogeneous rather than uniform.

Independently of geodetic coupling, a background-seismicity analysis using the Seismogenic Patches Detection (SPAD) algorithm [49] identifies seven persistent seismogenic patches (170–270 km along strike) spanning the Kamchatka segment of the Kuril-Kamchatka arc; these are interpreted as the seismic expression of tectonic asperities. The slip distribution of the 2025 rupture shows three separate maxima, each localized within a different patch and close to its margins [49]. This is consistent with rupture nucleating at asperity boundaries, and with the general expectation that stress concentrates at transitions between mechanically strong and weak fault-zone material [50]. Periollat and Funning [26] report the same boundary-nucleation pattern from an independent geodetic dataset: the $M_W$7.0 and $M_W$7.4 foreshocks of 2024–2025 are located near the edges, rather than the center, of the principal locked cluster, which they attribute to elevated stressing rates at asperity margins [51].

A third, independent line of evidence comes from gravimetry. Bürgmann et al. [48] report a broad correlation between negative offshore isostatic gravity anomalies and strongly coupled portions of the interface. They attribute more weakly coupled sectors—coincident with positive anomalies near two peninsulas—to the subduction of seafloor relief associated with the Emperor-Meiji seamount chain and the Kruzenstern fracture zone. Periollat and Funning [26] extend this analysis using trench-parallel gravity anomalies (TPGA) and show that the northern limits of both the 1952 and 2025 ruptures lie near a pronounced positive anomaly. They also show that the 2025 rupture nucleated near the transition between this positive anomaly and an adjacent negative anomaly, consistent with the interpretation that along-strike gravity structure marks persistent mechanical boundaries between segments rather than transient features of a single earthquake cycle.

Together, these three independent methods—background-seismicity clustering, probabilistic geodetic locking, and gravimetry—converge on a common picture: the segment that ruptured in 2025 is not a single homogeneous asperity but a cluster of at least three mechanically distinct subpatches bounded by persistent structural discontinuities. Within this cluster, rupture nucleation is systematically associated with patch margins rather than patch interiors.

A total of 19 $M_W \geq 5.9$ earthquakes occurred in the region (Fig. 7). They form two spatially and temporally distinct clusters. The first ("Group A") comprises nine earthquakes between 2014 and 2021, located ~48–51.5°N, near Severo-Kurilsk, and is dominated by a single $M_W$7.5 event on 25 March 2020. This event, classified as an intraplate, near-trench compressional event within the bending Pacific Plate [32], accounts for approximately 99% of the group's cumulative seismic moment release. The seismic activity in the region of interest shifted progressively northward in the following years. The second ("Group B") comprises ten earthquakes confined to a nine-month interval from August 2024 to July 2025 at ~52.4–53.0°N, directly overlapping the epicenter of the 2025 $M_W$8.8; all are consistent with interplate reverse faulting on the subduction interface. A similar spatial segmentation is also visible in the epicentral distribution of earthquakes from the complete portion of the catalog (Fig. 4, M ≥ 3.6).

No $M_W \geq 5.9$ earthquake occurred within 150 km of the eventual epicenter of the great *MW8*.8 earthquake between 2014 and August 2024. Given this absence, the cycles of the *dc-SSD* trajectory during 2014–2022 cannot correspond to the segment that subsequently hosted the 2025 rupture. Instead, we attribute them to the Group A source region and to other, more distant structures along the arc, each undergoing an independent, complete localization cycle on its own patch. This is consistent with Ostapchuk and Nugmanov [49], who report that individual asperities along the Kamchatka segment can rupture repeatedly before eventually rupturing jointly with their neighbors.

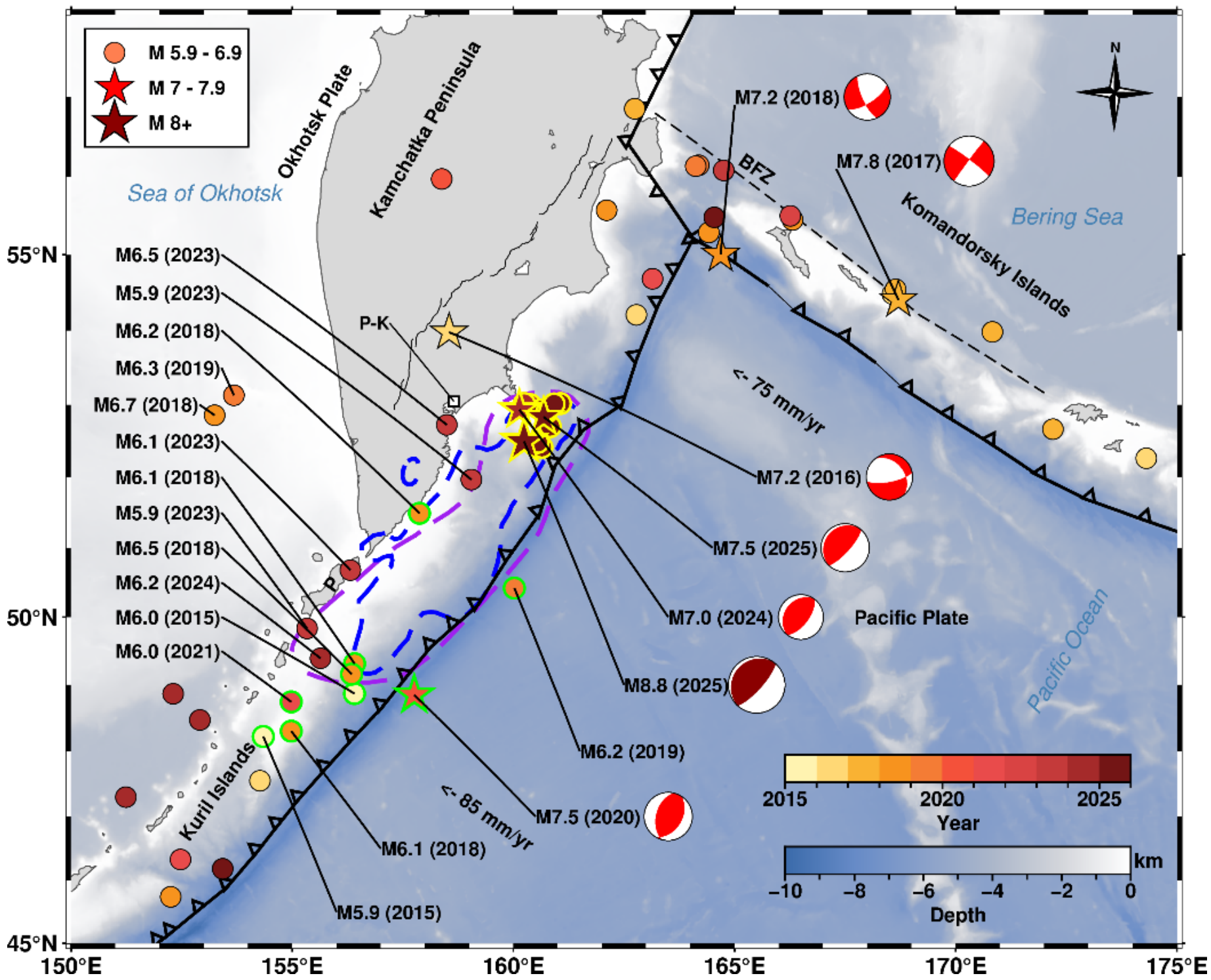

*Figure 7. Earthquakes of $M_W \geq 5.9$ in the Kamchatka region, 20 July 2014–29 July 2025. Marker color indicates time of occurrence (light yellow to dark red, oldest to most recent) and marker type scales with magnitude. Green outlines mark Group A (nine events, 2014–2021, ~48–51.5°N, near Severo-Kurilsk), dominated by the $M_W$7.5 event of 25 March 2020; yellow outlines mark Group B (ten events, August 2024–July 2025, ~52.4–53.0°N), directly overlapping the epicenter of the 29 July 2025 $M_W$8.8 mainshock (labeled). Major and minor tectonic features from the GEM Global Active Faults Database [62] are represented by thick and thin black lines, respectively. Arrows indicate the direction and velocity of motion of the Pacific Plate relative to the Okhotsk plate [19]. Active volcanoes from the Global Volcanism Program [63] are represented by triangles, with Krasheninnikov volcano highlighted. The area that experienced at least 2 m of slip during the 2025 $M_W$8.8 rupture is outlined by a blue dashed line based on Liu et al. [1], with adjacent $M_W$5.9+ earthquakes labeled. The purple outline represents the rupture area of the $M_W$9.0 1952 earthquake from MacInnes et al. [25]. A thin dashed line represents the Bering Fracture Zone (BFZ). The city of Petropavlovsk-Kamchatsky is marked by a square (P-K). P marks Paramushir Island. The labels of the $M_W$7+ earthquakes are accompanied by their moment tensor solutions from GCMT [34,35], with red used for magnitudes from 7 to 8 and dark red for magnitudes above 8.*

The regional *dc* series reaches a well-defined minimum in early 2022, approximately two years after the dominant Group A event ($M_W$7.5, March 2020, Episode 3), and about half a year after the smaller earthquake from this group ($M_W$6.0, August 2021, Episode 4). We interpret this period from the mentioned events to the *dc* minimum as most consistent with viscoelastic stress transfer from the Group A source region into the still-locked segment that later hosted the 2025 rupture. Coseismic Coulomb stress transfer from the same source may also have contributed, but at the ~150–250 km separation, its effect would be expected to account for a smaller fraction of the total perturbation. Postseismic viscoelastic relaxation in subduction zones typically proceeds over Maxwell timescales on the order of 10–25 years. This estimate is based on asthenospheric viscosities of 2.3–3.9×$10^{19}$ Pa·s fitted directly to decades of postseismic GPS and tide-gauge data following the 1964 $M_W$9.2 Alaska earthquake; such values are regarded as broadly typical of subduction zones [52]. On this timescale, the 2022 *dc* minimum, occurring two years after the major Group A event, corresponds only to the early, rising portion of a relaxation process that would not be expected to complete for a decade or more—consistent with a real but still-developing viscoelastic contribution. We note that Vladimirova et al. [53] report a substantially lower Maxwell viscosity (3×$10^{17}$ Pa·s) for the segment of the Kuril arc immediately adjacent to the 2006–2007 Simushir rupture, which they attribute to local rheological weakening in that specific zone. We do not use this lower value here because there is no equivalent evidence for anomalous weakening beneath the Group A–Group B segment, and the more general subduction-zone range provides the more appropriate comparison.

The magnitude of stress change expected from such a mechanism is small relative to near-field coseismic values but not negligible when accumulated over years. For a comparably sized ($M_W$7.7) strike-slip earthquake, Li et al. [54] computed static Coulomb stress changes of up to 3 kPa on fault segments several hundred kilometers away, increasing to ~5.5 kPa after 20 years of postseismic viscoelastic relaxation—below, but approaching, the ~10 kPa (0.1 bar) stress change commonly regarded as sufficient to promote earthquake triggering [55]. We use this as an order-of-magnitude analogue rather than a Kamchatka-specific estimate: no dedicated stress-transfer calculation was performed for the Group A–Group B pair in this study.

We propose that this *dc* minimum required the joint satisfaction of four conditions. (i) The receiving segment was already close to a critical stress state after decades of essentially complete interseismic locking [48], so its own internal production of stress heterogeneity, $Q_{\tau,int}$, was low. (ii) The incoming stress perturbation was spatially coherent: viscoelastic relaxation produces a smoothly varying stress field over hundreds of kilometers, in contrast to the sharply alternating, short-wavelength pattern of stress increase and decrease produced by a single coseismic Coulomb stress change. (iii) Sufficient time had elapsed for this perturbation to propagate into the segment: viscoelastic stress transfer builds up progressively over the Maxwell relaxation timescale as the ductile lower crust and upper mantle flow in response to the coseismic load. (iv) The segment's structural boundaries, independently marked by gravity-anomaly transitions [26,48], confined the perturbation within the segment: contrasts in effective rigidity across these boundaries are expected to impede along-strike stress transmission, so stress entering from the south tended to remain within the segment rather than dissipating further along the arc. Together, these four conditions explain why, in the first months of 2022, the balance between the production of stress heterogeneity ($Q_\tau$) and its damping through χ(D)-mediated redistribution shifted in favor of damping. The arrival of coherent external stress from the Group A region, delayed by ~2 years, could provide the first synchronizing influence. Because internal production was already low (condition (i)) and external redistribution acted coherently, had sufficient time to propagate, and remained confined within the segment (conditions (ii)–(iv)), the variance of the inelastic strain-rate field, represented by *dc*, was driven toward the minimum.

Following the early 2022 minimum, *SSD* rises to a local maximum around mid-2023. Because the mean rate of inelastic deformation $\langle \dot{\gamma}_{in} \rangle$ increases monotonically as the damage-softened viscosity $\eta(D)$ declines (Equation 4), we interpret this *SSD* rise as evidence of continued damage accumulation within the newly localizing zone. *SSD* subsequently declines to its lowest value in the record (August 2024). The *dc* minimum in May–August 2024, unlike those in the shorter cycles of 2014–2022, is elevated relative to the 2022 *dc* minimum, with a value around the 50% benchmark line (Figs. 5a and 6a). Thereafter, *dc* rises through mid-2025, with a shallow decline preceding the seventh episode of strong earthquakes, which includes the $M_W$8.8 event (Fig. 6a). This pairing of declining *SSD* with persistently elevated *dc* is consistent with the behavior associated with Stage IV of the five-stage damage model [11], in which *dc* recovers as rupture-scale complexity develops while *SSD* plateaus or declines. The absence of a renewed deep *dc* minimum after 2022 is therefore consistent with the growing complexity of the localizing deformation as it approached rupture, rather than a return to the distributed, multi-site deformation of the earlier cycles.

The $M_W$7.0 "Shipunsky" earthquake of 17 August 2024 [56,57] is the first interplate rupture in the decade-long record to occur within 150 km of the eventual $M_W$8.8 mainshock, and it coincides with the onset of Group B activity. Static stress-transfer modelling by Stein et al. [58] indicates that this event brought the future epicentral patch of the $M_W$7.5 foreshock rupture approximately 1.5 bar closer to failure, while reducing stress on the northwest corner of the eventual $M_W$8.8 mainshock rupture surface. Independent support for this type of cascading stress transfer along the same rupture sequence comes from Xiao et al. [59], who compute Coulomb failure stress changes (ΔCFS) imparted by the $M_W$7.5 foreshock to the mainshock fault and by the mainshock to its subsequent $M_W$7.4 and $M_W$7.8 aftershocks, confirming a consistent triggering cascade through the sequence. The $M_W$7.5 foreshock of 20 July 2025 was itself accompanied by afterslip that reportedly drove the expansion of its aftershock sequence and contributed to triggering the nucleation of the mainshock nine days later [59]. This sequence—a moderate event on the eventual rupture segment (August 2024), followed by afterslip-driven promotion of nucleation on the adjacent locked patch (July 2025 foreshock), and culminating in rupture of the full segment nine days later—matches the "delayed cascade-up" pathway identified in rate-and-state simulations of hierarchical asperity systems. In this pathway, a small-patch earthquake's afterslip migrates into the surrounding locked patch and attracts large-scale nucleation to its location, with the eventual large event initiated by re-acceleration at the margin of the precursory rupture [60]. The $M_W$8.8 mainshock of 29 July 2025 ruptured three separate patches within a single continuous rupture process [49]. The spatially mixed, patch-by-patch ΔCFS pattern imparted by the August 2024 event and the July 2025 foreshock [58,59] added a further short-lived episode of stress heterogeneity in the final weeks before the mainshock, superimposed on the longer maturation phase.

We therefore conclude that the *SSD*-*dc* record spanning 2014–2025 does not describe a single, continuous preparatory sequence for the 2025 $M_W$8.8 earthquake. Instead, it records two processes operating at different spatial and temporal scales and superimposed within the same regional catalog. Between 2014 and 2022, shorter *SSD*-*dc* cycles reflect the independent, complete localization histories of structures elsewhere along the arc—principally Group A, but also more distant segments—each following the isolated-fault behavior described above and not contributing to the segment that later ruptured in 2025. These cycles recur with *dc*-minimum-to-event intervals ranging from eight days to one year, with each cycle closing before the next *dc* minimum begins. Superimposed on these cycles is a second, slower process specific to the eventual rupture segment: a single *dc* minimum in the first months of 2022, followed by an interval of sustained, non-recovering *dc* behavior through to the mainshock, exhibiting only shallow minima. From this minimum at the beginning of 2022 onwards, the earlier recurrence pattern changed: the same minimum was followed first by the August 2024 Shipunsky earthquake during Episode 6 and then, after a shallow intervening *dc* minimum, by the July 2025 mainshock.

The three-patch structure invoked to interpret the 2025 rupture was established independently of the *dc*-*SSD* analysis, using background-seismicity clustering [49], probabilistic geodetic locking [26,48], and gravimetry [26,48]. All three methods place the same structural boundaries along the segment, and all three associate rupture nucleation with these boundaries rather than with patch interiors. The *SSD*-*dc* trajectory adds a fourth, independent line of evidence to this convergence. It is obtained from a different type of observation and provides temporal resolution that the other three methods lack. Its principal contribution is therefore not the location of the segment boundaries, which was already established, but the timing of the transition between the two regimes described above—in early 2022, when the segment began behaving as a coupled system rather than as three independent patches.

The central methodological claim of this study is that *dc* and *SSD*, computed from catalog seismicity alone, tracked a physical process whose independent signature is also visible in geodetic, seismological, and gravimetric data collected by entirely different means. This claim rests on three specific points of agreement. First, the theoretically predicted *dc*-minimum-before-*SSD*-maximum sequence (Orlecka-Sikora et al., GJI-26-0445.R1) occurred in every cycle identified in the record, including the final one, despite differences among the cycles in duration, amplitude, and triggering mechanism. Second, the timing of the early 2022 *dc* minimum is consistent, within the available precision, with an independently estimated propagation time for stress transfer from the Group A source region. Third, the three patches identified independently by SPAD ruptured together, within a single rupture process, in the same earthquake whose approach is recorded as a sustained, non-recovering *dc* elevation.

This distinction matters in light of an independent analysis of the same rupture zone based on seismicity rate. Stein et al. [58] show that the rate of $M_W \geq 4.5$ earthquakes over the future 2025 rupture area was steady from 1995, when the catalog becomes complete at this magnitude, until essentially the mainshock itself, with neither long-term slowdown nor acceleration, and conclude that the seismicity rate at this location is not indicative of precursory activity until the final ten days before the earthquake. *SSD* and *dc* are both derived from the same underlying seismicity as this rate analysis, yet the *dc* trajectory identifies a specific, dated transition—the early 2022 *dc* minimum together with an increase in SSD—about three years before the mainshock. Furthermore, the *dc* exhibited two other shallow but distinct minima, approximately three months before the "Shipunsky" earthquake in the sixth episode and five months before the $M_W$8.8 mainshock, respectively.

The *dc* is sensitive to changes in the spatial and temporal structure of deformation that leave no trace in the rate of moderate seismicity. As *dc* rose from its 2022 minimum toward its subsequent high in October 2023, *SSD* rose in parallel before reaching a plateau around mid-2023—a pattern consistent with rheological saturation rather than with a reset of the localization cycle. The subsequent declines in *dc* through late 2023–early 2024 and late 2024–early 2025 are temporally associated with the ruptures of substantial fragments of the eventual rupture zone—the $M_W$7.0 Shipunsky earthquake (August 2024) and the $M_W$7.5 foreshock (July 2025).

Several limitations should be kept in mind when interpreting the results above. The viscoelastic relaxation mechanism proposed for the early 2022 *dc* minimum is supported only as an order-of-magnitude consistency check against published viscosity estimates for a neighboring segment; no dedicated stress-transfer model was constructed for the present segment. The spatial correspondence between SPAD patches, locking subclusters, and gravity-anomaly boundaries is qualitative and is based on a visual comparison of published maps rather than on shared coordinates for the underlying datasets. Finally, the Group A/Group B classification used here is based on fixed geographic bounding boxes rather than on a formal patch-assignment procedure; although it is consistent with the independently derived patch geometry, it was not derived directly from that geometry.

## Data availability

The earthquake catalogue data analyzed in this study are publicly available from the International Seismological Centre On-line Bulletin [33]. The processed catalogues and dc/SSD time series generated during the current study are available from the corresponding author on reasonable request.

## Code availability

The custom MATLAB codes used to calculate the seismic regularity coefficient (dc), seismic strain dynamics (SSD), and their temporal evolution are available in the IG PAS Data Portal [64]. The repository includes the computational routines and documentation describing the analysis workflow.

## Acknowledgements

Figures 1, 2 and 4 were created with Generic Mapping Tools (GMT), version 5 [61].

## Funding

This work was supported by the National Science Centre, Poland, Grant No. OPUS 2024/55/B/ST10/01041, Agreement No. UMO-2024/55/B/ST10/01041. The research used infrastructure co-financed by the Ministry of Science and Higher Education under Contract No. 2024/WK/05.

## Author contributions

Conceptualization: S.L., B.O-S.; Data curation: M.O.; Formal Analysis: S.L., M.O.; Funding acquisition: B.O-S.; Investigation: B.O-S., S.L., M.O.; Methodology: B.O-S., S.L.; Software: S.L.; Visualization: M.O., S.L.; Writing – original draft: M.O., B.O-S., S.L.; Writing – review & editing: S.L.

## Competing interests

The authors declare no competing interests.